%% file: camsap.tex
\documentclass{article}
\usepackage{standalone}

\usepackage{style/spconf}
\usepackage{amsmath, amsthm, amssymb}
\usepackage{bm}
\usepackage{dsfont}
\usepackage{graphicx}

\usepackage{cite} % Orders citations.
\usepackage{url}
\usepackage{hyperref}

\usepackage{tikz}
\usetikzlibrary{positioning}
\usetikzlibrary{calc}

\usepackage{style/macros}

\newtheorem{theorem}{Theorem}
\newtheorem{assumption}{Assumption}

\newcommand{\inW}{A}
\newcommand{\outW}{B}
\newcommand{\Linf}{\calL}
\newcommand{\Lfin}{\calL_\sqcap}
\newcommand{\avg}[1]{ \lim_{w \to \infty} \frac{1}{w} \int_{\calC_w} #1 d\bfp}

\title{Multi-Target Tracking with Transferable Convolutional Neural Networks}
\name{
\begin{tabular}{@{}c@{}}
Damian Owerko${}^{\star}$ \qquad Charilaos I. Kanatsoulis${}^{\star}$ \qquad Jennifer Bondarchuk${}^{\dagger}$ \\
Donald J. Bucci Jr${}^{\dagger}$\qquad Alejandro Ribeiro${}^{\star}$
\end{tabular}
}
\address{
${}^{\star}$ Department of Electrical and Systems Engineering, University of Pennsylvania, PA, USA \\
${}^{\dagger}$ Advanced Technology Labs, Lockheed Martin, Cherry Hill, USA
}

\begin{document}
\ninept
\maketitle
\begin{abstract}
    Multi-target tracking (MTT) is a classical signal processing task, where the goal is to estimate the states of an unknown number of moving targets from noisy sensor measurements. In this paper, we revisit MTT from a deep learning perspective and propose a convolutional neural network (CNN) architecture to tackle it. We represent the target states and sensor measurements as images and recast the problem as an image-to-image prediction task. Then we train a fully convolutional model at small tracking areas and transfer it to much larger areas with numerous targets and sensors. This transfer learning approach enables MTT at a large scale and is also theoretically supported by our novel analysis that bounds the generalization error. In practice, the proposed transferable CNN architecture outperforms random finite set filters on the MTT task with 10 targets and transfers without re-training to a larger MTT task with 250 targets with a 29\% performance improvement.
\end{abstract}
\section{Introduction}
\label{sec:intro}

Multi-target tracking (MTT) involves estimating the state of multiple moving targets from noisy sensor data \cite{Vo15-MultitargetTracking}. MTT is an important task and was originally introduced for aerospace surveillance, e.g., to track airplanes using radar for air traffic control \cite{bar1974extension,reid1979algorithm}. Recently MTT approaches have been used in various other fields including but not limited to robotics \cite{hu2012cloud,ferri2017cooperative}, health analytics \cite{genovesio2006multiple,mavska2014benchmark}, and autonomous driving \cite{patole2017automotive}.

A core challenge of the MTT problem deals with properly handling \emph{measurement origin uncertainty} (MOU). In particular, it is unknown whether measurements come from targets or clutter and whether or not existing targets are miss-detected or have disappeared. Joint probabilistic data association (JPDA) filters handle MOU by creating posterior mixture distributions per target, based on marginal association probabilities. Multiple hypothesis tracking (MHT) approaches, on the other hand, maintain multiple data association decisions per target over a short history. Other methods use global nearest neighbor tracking, which is a memoryless version of MHT, and is used in many practical systems.

% global nearest neighbor (GNN) tracking, and random finite set (RFS) \cite{Mahler07-Statistical, Mahler14-AdvancesStatistical} based approaches. 
% In JPDA the core methodology is making soft association decisions by creating a posterior mixture distribution per target based on the marginal association probabilities. MHT, on the other hand, makes deferred hard association decisions about what measurements likely go with which targets by maintaining consistency over a short history.
%  GNN methods are basically memoryless MHTs basically and are usually implemented in practical systems. \red{Why don't we talk more about these approaches?}

% RFS approaches provide a Bayesian analysis of the entire MTT problem without an explicit association model/distribution. In particular,

Random finite sets (RFS) \cite{Mahler07-Statistical, Mahler14-AdvancesStatistical} is a large class of MTT approaches that provide a Bayesian analysis of the MTT problem and have shown remarkable MTT performance.
They were first popularized for unlabeled multi-object state estimation via the Probability Hypothesis Density (PHD) \cite{Vo06-GaussianMixture,vo2003sequential} and Cardinalized Probability Hypothesis density (CPHD) filters \cite{vo2007analytic}.
Labeled RFS filters \cite{Vo13-LabeledRandomFinite} were introduced to better approximate the optimal Bayesian filter. The Generalized Labeled Multi-Bernoulli (GLMB) filter \cite{do2019tracking} models the target states as a mixture of hypotheses, each representing a possible combination of target labels. Each hypothesis contains mutliple single-target probability densities.
The Labeled Multi-Bernoulli (LMB) filter \cite{Reuter2014} reduces the complexity of the GLMB. It uses a single mixture where each component represents a possible measurment-target association.
The LMB and GLMB filters provide excellent target tracking performance, but widespread adoption is constrained by computational scalability, especially with multiple sensors. Efficient computing approaches for these filters are an area of active reserach \cite{Vo17-EfficientImplementation, Vo19-MultiSensorMultiObject,beard2020solution}.

Deep learning approaches promise to alleviate computational complexity problems. Transformer-based models have comparable performance to RFS fitlers on simple problems and provide state-of-the-art results on more complex tasks \cite{Pinto21-NextGeneration, Pinto22-CanDeepLearning}. However, their scalability is limited by quadratic runtime with respect to the number of targets and measurements \cite{Vaswani17-AttentionAllYou}.

In this paper, we propose a novel transfer learning approach to perform large-scale MTT with \emph{multiple sensors}. In particular, we recast the MTT problem as an image-to-image prediction task and train a convolutional neural network (CNN) model to perform MTT. To overcome the scaling limitation, the CNN model is trained on small windows of the tracking area, but executed on much larger areas. The runtime of the model is proportional to the tracking area, enabling MTT at previously intractable scales. The approach is supported by theoretical analysis that provides a bound on the generalization performance of CNNs to large signals. Numerical experiments on several MTT scenarios showcase the effectiveness of the proposed approach. In particular, we show that training can be performed in $1\text{km}^2$ window and the performance does not degrade as we scale up. Instead it improves by $29\%$ on a $25\text{km}^2$ window.

% Recent works have shown that CNNs trained on small scale spatial problems can transfer their performance to larger problems without retraining \cite{Mox22-Learning}.

% would think you want to claim that this transformer approach is not great because the dimensionality of the network is linked to the number of targets/measurements. The issue of guaranteeing transferrability is also problematic, because it means you have to train of very large data sets which could take a long time.

\section{Problem Overview}\label{sec:mtt}

Let $\bfx_{n,i} \in X$ be a vector that represents the state of the $i$-th target at time $n$ in some state space $X$. In numerous applications, the state of each target is a vector $\bfx_{n,i} \in \reals^4$ of positions and velocities given by equation \eqref{eq:state}.
\begin{equation}\label{eq:state}
    \bfx_{n,i} = \bmat{\bfp_{n,i} & \bfv_{n,i}}
\end{equation}
where $\bfp_{n,i},~\bfv_{n,i} \in \reals^2$ are the two-dimensional position and velocity respectively.
Then, the \emph{multi-target state} at time $k$ is a set $\calX_n = \{\bfx_{n,1}, ..., \bfx_{n,|\calX_n|} \}$, where $|\calX_n|$ is the possibly time varying cardinality of the multi-target state.
Between each time step the multi-target state can change in three ways. First, the states of individual targets evolve according to an unknown model. Second, existing targets may die with probability $p_{\text{death}}$. Third, new targets may be born according to a Poisson distribution with mean $\lambda_{\text{birth}}$. The goal of MTT is to estimate \( \calX_n \) from noisy measurements thereof.

In an MTT environment, there are $M$ sensors, which make observations of the targets in some space $Z$. At each time step a sensor $s \in \{1,...,M\}$  can detect each target $\bfx \in \calX_n$ with probability $p_{D}$. An observation for a detected target $\bfz^s \in Z$ is distributed according to a density $g_D^s(\bfz^s | \bfx_{n,i})$. The sensor may make false-positive observations, called \emph{clutter}.
% The number of clutter returns per sensor is Poisson distributed with rate $\lambda^s_C$, and the distribution of clutter measurements is assumed to be uniformly distributed.
Let $\calZ^s_n$ be the set of all true observations and clutter, made by the sensor at time $n$, so that $\calZ_n = \cup_{s \in \{1,...,M\}} \calZ^s_n$ is the set of observations for all sensors. We denote the sequence of past observations as $\calZ_{1:n} = (\calZ_1, ,...,\calZ_{n})$.

\section{Convolutional Neural Networks for Multi-Target Tracking}\label{sec:cnn}
As mentioned in the introduction state-of-the-art MTT approaches suffer from computational complexity issues, especially when the number of targets grows large. To overcome these limitations, we propose a novel CNN approach that exploits the spatial symmetries of MTT, by leveraging the shift equivariance property of CNNs \cite{cohen2016group}. A CNN is described by the following recursive equation:
\begin{equation}\label{eq:cnn}
    x_l(\bfp) = \sigma \left( \int_{\reals^2} h_l(\bfs) x_{l-1}(\bfp - \bfs) d\bfs \right)
\end{equation}
where \(x_l\) is the output at the \(l^\text{th}\) CNN layer, which is produced by the convolution of the input $x_{l-1}$ with the filter \(h_{l-1}\), followed by a pointwise nonlinearity \(\sigma_l\).

Our approach casts MTT as an image-to-image task. To that end, we predict the positions of each target, \( \bfp_{n,i} \). The sensor measurements are noisy observations of these positions. We represent the observations and the positions by two-dimensional signals, and perform MTT by learning a CNN that maps the observations image to the positions image.
The MTT multi-object state, $\calX_n$ is represented by a 2D intensity function. Specifically, we construct a superposition of Gaussian pulses with variance $\sigma^2$, centered at the position of each of the targets. Equation \eqref{eq:intensity-target} is the target intensity function at a point $\bfp \in \reals^2$.
\begin{equation}\label{eq:intensity-target}
    U(\bfp; \calX_n ) = \sum_{i \in N_n} (2 \pi \sigma)^{-1}
    \exp(-\frac{1}{2\sigma^2}||\bfp-\bfp_{n,i}||^2_2)
\end{equation}
The target intensity function \eqref{eq:intensity-target} encodes target positions as a 2D image. A CNN predicting this intensity function $U$ is analogous to learning a PHD filter using a CNN.

To represent the set of sensor measurements \( \calZ^s_n \) as an intensity function we assume the existance of a sensor measurement model, \( g_D^s(\bfz^s | \bfx_{n,i}) = g_D^s(\bfz^s | \bfp_{n,i}) \). We construct an intensity function \(V(\bfp; \calZ_n)\) as a superposition of the densities \( g_D^s \):
\begin{equation}\label{eq:intensity_observation}
    V(\bfp; \calZ_n) = \sum_{\bfz^s \in \calZ_n} g_D^s(\bfz^s | \bfp).
\end{equation}
Note that in \eqref{eq:intensity_observation} the set of observations \(\calZ^s_n\) includes both true detections and clutter.

Our approach is to learn a CNN, with parameters $\calH$, that maps $U$ to $V$, by minimizing their expected average squared error:
\begin{equation}\label{eq:mse_inf}
    \Linf(\calH) = \E{ \avg{ |\bm\Phi(V; \calH)(\bfp) - U(\bfp)|^2 } }.
\end{equation}
In Equation \eqref{eq:mse_inf}, \( C_w \) is an open set of the points in \( \reals^2 \) within a square of side lengths \( w \), centered around the origin.
In practice, using \eqref{eq:mse_inf} for CNN training is intractable for two reasons. First, it requires data processing over all real numbers. Even though typical MTT signals have finite support, at large scales they are too wide to efficiently evaluate \eqref{eq:mse_inf}. Second, the intensity functions are continuous. The first challenge is addressed in Section \ref{sec:window} and the second in Section \ref{sec:architecture}.

\section{A Transfer Learning approach to Multi-target tracking}\label{sec:window}
To overcome the scalability challenge, we propose a transfer learning approach, where we train the CNN with small windows of the tracking area and apply the trained CNN to much larger areas with performance guarantees. Instead of minimizing \eqref{eq:mse_inf} directly, we solve a smaller problem with windowed versions of the input-output signals \(V,~U\). In particular, we consider windows over the input and output signals of width \(\inW\) and \(\outW\), when \(\inW \ge \outW\). Then the problem in \eqref{eq:mse_inf} can be cast as:
\begin{equation}\label{eq:mse_window}
    \Lfin(\calH) = \frac{1}{\outW^2}
    \E{ \int_{\calC_\outW} | \bm\Phi(\sqcap_\inW V; \calH)(\bfp) - U(\bfp) |^2  d\bfp },
\end{equation}
where \(\sqcap_\inW\) is the indicator function $\sqcap_\inW(\bfp) \defeq \mathds{1}(\bfp \in \calC_\inW)$.
The problem in \eqref{eq:mse_window} can be easily optimized since the signals involved are space-limited, especially when \(\inW, \outW\) are small. Then the optimized CNN can be executed in larger windows, thus enabling MTT at previously infeasibly large areas with high accuracy. Next, we provide a theoretical analysis that quantifies the performance degradation when training on a small window, but executing on the whole area, by deriving an upper bound for \(\Linf\) in \eqref{eq:mse_inf} in terms of the cost \(\Lfin(\calH)\) in \eqref{eq:mse_window}. To do so we model the state and observation intensity functions as stochastic processes that satisfy the following assumption.

\begin{assumption}\label{assume:jointly_stationary}
    \(U\) and \(V\) are jointly stationary and bounded so that \( |U(\bfp)| < \infty \) and \( |V(\bfp)| < \infty \) for all \(\bfp \in \reals^2\).
\end{assumption}

\noindent This is the main assumption of our analysis and is crucial for exploring the transferability properties of the CNN in performing MTT on a large scale. In practice, we observe that the state and observation intensity functions exhibit approximate stationarity, which supports the applicability of our approach.

Our analysis studies the output of a CNN when the input is stationary. To do so we make the following assumptions, that are satisfied by a typical CNN implementation.

% Let \(\bm\Phi(X; \calH)(t) := x_L(t) \) be the output of a CNN with \(L\) layers, input \(x_0(t) := X(t)\), and a set of filters \(\calH = \{ h_1,...,h_L\}\) as described by \eqref{eq:cnn}. Then, the expected average squared error between the stationary input and output of the CNN takes the form.

\begin{assumption}\label{assume:finite_filters}
    The filter functions \( h_l \in \calH \) are continuous with a finite width \(K\). That is \(h_l(\bfp) = 0\) for all \(\bfp \notin [-K/2, K/2]^2\).
\end{assumption}
\begin{assumption}\label{assume:filters_bounded}
    The filters have finite L1 norms \(|| h_l ||_1 \le \infty\) for all \( h_l \in \calH \).
\end{assumption}

\begin{assumption}\label{assume:lipschitz}
    The nonlinearities \( \sigma_l(\cdot) \) are normalized Lipshitz continuous (the Lipshitz constant is equal to 1).
\end{assumption}
\noindent The majority of pointwise nonlinear functions used in deep learning, e.g., ReLU, Leaky ReLU, hyperbolic tangent, are normalized Lipshitz for numerical stability.

Theorem \ref{thm:stationary_bound} quantifies the difference between the mean squared error of a CNN trained on a finite window according to \eqref{eq:mse_window} and executed on an infinite window of the tracking area.
\begin{theorem}\label{thm:stationary_bound}
    Consider a CNN, defined in \eqref{eq:cnn}, with \(L\) layers and a set of filter parameters \(\hat\calH\). Let \(\Lfin(\hat\calH)\) be the cost the CNN achieves on the windowed problem as defined by \eqref{eq:mse_window} with an input window of width \(\inW\) and an output window width \( \outW \). Under Assumptions \ref{assume:jointly_stationary}-\ref{assume:lipschitz} the associated cost \( \Linf(\hat\calH) \) on the infinite problem, as defined by \eqref{eq:mse_inf}, is bounded by the following.
    \begin{align}
         & \Linf(\hat\calH) \le \Lfin(\hat\calH) + \E{X^2} C + \sqrt{\Lfin(\hat\calH) \E{X^2} C} \label{eq:stationary_bound} \\
         & C = \frac{H^2}{\outW^2} \max(0,(\outW + LK)^2 - \inW^2),~ H = \prod_l^L || h_l ||_1
    \end{align}
\end{theorem}
The proof is relegated to the journal version of the paper. The key idea of the proof leverages the shift-equivariance property of CNNs in order to exploit the spatial properties and stationarity of the two-dimesnional input-output signals.  Essentially, Theorem \ref*{thm:stationary_bound} demonstrates that minimizing \( \Linf \) can be approximated by minimizing \( \Lfin \). The potential performance decrease is influenced by the variance of the input signal, the number of layers, filter widths, and the sizes of the input and output windows. The second and third term on the right-hand side of \eqref{eq:stationary_bound}, appear due to padding and can be eliminated when
\( \inW \ge \outW + LK \).

% The inequality presented in \eqref{eq:stationary_bound} simplifies to 
% \(\Linf(\hat\calH) \le \Lfin(\hat\calH)\). In this particular scenario, the padding has no impact on the output, which is why the second term appears on the right-hand side of \eqref{eq:stationary_bound}.

%  The maximum deterioration in performance is bounded by a quantity that is affected by the variance of the input signal, number of layers, filter widths, and the size of the input and output windows. The inequality in \eqref{eq:stationary_bound} reduces to \(\Linf(\hat\calH) \le \Lfin(\hat\calH)\) whenever \( \inW = \outW + LK \). In this special case the output becomes unaffected by padding, which explains the second term in the right hand side of \eqref{eq:stationary_bound}.

% You should really focus here on highlighting the intuitiveness of this result. We are basically saying that for this type of CNN architecture, the training error achieved on the windowed signal versus the full signal is bounded by the signal content lost as a result of the windowing.

\section{Proposed Architecture}\label{sec:architecture}
To implement our proposed approach, we train a CNN, formed by discrete convolutions on discrete signals. To do so, we sample the intensity functions at regular intervals, i.e.,
\begin{align}
    \bfU_{ij}(\calX_n) & = U( \rho\bmat{i & j}; \calX_n ) \label{eq:image_targets}       \\
    \bfV_{ij}(\calZ_n) & = V( \rho\bmat{i & j}; \calZ_n ) \label{eq:image_observations},
\end{align}
The scalar $\rho$ is the spatial resolution (units of meters per pixel) at which we sample. $\bfU, \bfV \in \reals^{N \times N}$ are the resulting images that describe the state $\calX_n$ and sensor observations $\calZ_n$, respectively. Note that the spatial resolution has to be sufficiently small to avoid aliasing.
In practice, we learn a CNN model $\bm\Phi$ which, given a sequence of $K$ past observation images
$\tilde{\bfV}(\calZ_{n-K+1},...,\calZ_n) = (\bfV^{n-K+1},...,\bfV^{n-1},\bfV^n)$
estimates the target image, $\bfU^n$.

To perform the MTT task we propose a model with a fully convolutional encoder-decoder architecture. A schematic illustration is presented in Figure \ref{fig:encoder_decoder}. The encoder and decoder layers perform downsampling and upsampling operations, respectively. In the following experiments, the encoder is formed by a stack of three layers, consisting of 2D convolutions with a stride of 2, kernel size of 9, and 128 channels. The decoder uses three convolutional layers with 2D transpose convolutions and the same hyper-parameters. In between the encoder and the decoder, there are four hidden layers with 2D convolutions, kernels of size 1, and 1024 channels. The CNN is built with leaky-ReLU activations with parameter 0.01.
\begin{figure}[ht]
    \centering
    \includestandalone[width=\linewidth]{figures/architecture_small}
    \caption{Diagram of the proposed CNN architecture. Above we depict a CNN with a single encoder, hidden, and decoder layer. The filter parameters \( \bfH_l \in \reals^{N \times N \times F_l \times F_{l-1}} \) are represented by \( \bfH_l[i,j] \in \reals^{F_l \times F_{l-1}} \), which is indexed by \(i,j \in \mathbb{Z}\). }
    \label{fig:encoder_decoder}
\end{figure}
\begin{figure*}
    \centering
    \includegraphics[width=\linewidth]{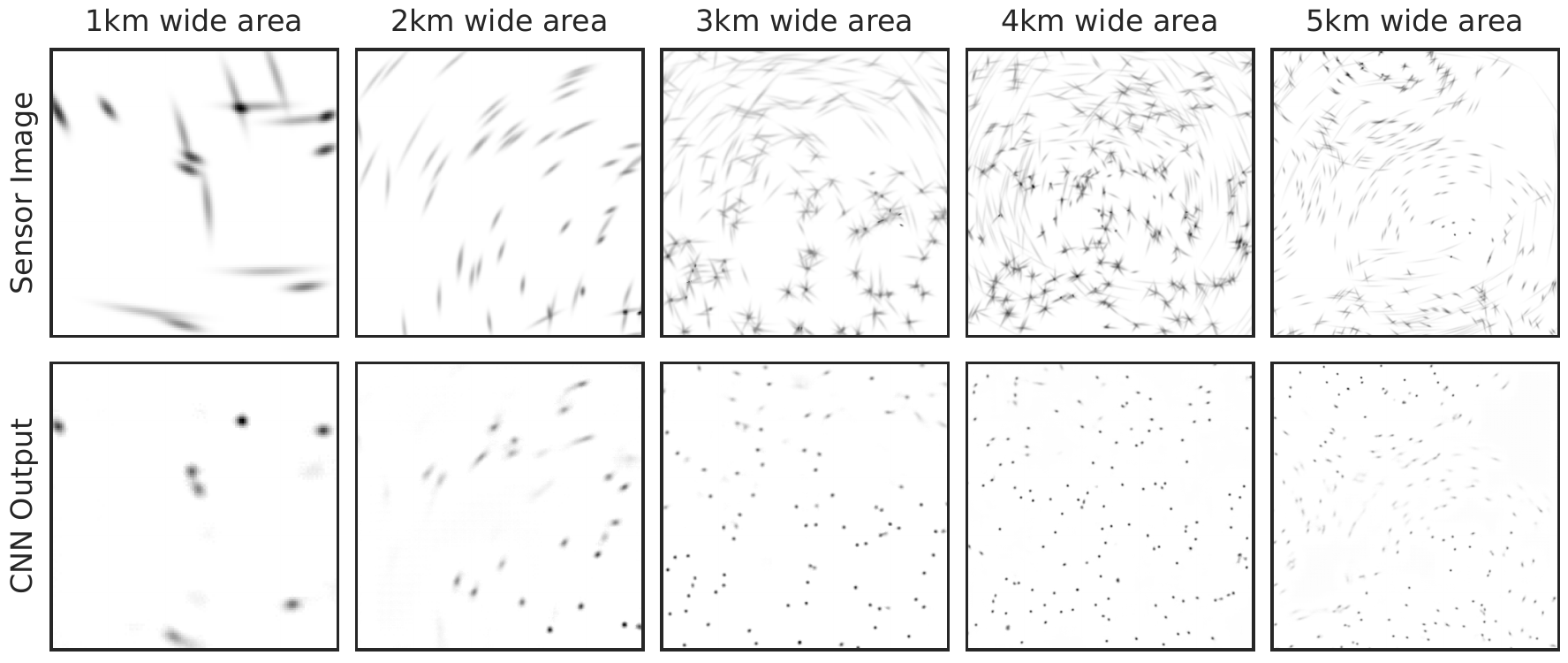}
    \caption{The input and output of the CNN at different window widths of \(w = \{1,...,5\} \) km. The first row shows the input sensor image, whereas the second row shows the corresponding output of the trained CNN.}
    \label{fig:multiscale}
\end{figure*}

\section{Numerical Experiments}\label{sec:experiments}

To simulate a MTT environment, we consider a square \emph{rendering window} of width $w$ km that represents the tracking area. The sensors can be located outside this window, but still detect targets within the boundaries of the window. Therefore, we simulate the targets and sensors on a larger concentric square \emph{simulation region} of width $W = w + 2R$ where $R$ is the maximum range of the sensors. The number of targets and sensors, at the beginning of each simulation, are drawn from $\Poisson(10 W^2)$ and $\Poisson(0.25W^2)$, respectively. Their initial positions are uniformly distributed within the simulation region.

The state $\bfx_{n,i}$ evolves according to the (nearly) constant velocity (CV) model \cite{RongLi03-SurveyManeuvering} given by
\begin{equation}\label{eq:cv}
    \begin{split}
        \bfp_{n+1,i} = \bfp_{n,i} + T \dot{\bfp}_{n,i} + \frac{T^2}{2} \eta_{n,i}
        \quad&\bfv_{n+1,i} = \bfv_{n,i} + T \eta_{n,i}.
    \end{split}
\end{equation}
The model is discretized with time-step $T = 1$ seconds. The acceleration $\eta_{n}\sim\calN(0,\sigma_\eta^2)$ normally distributed with $\sigma_\eta^2 = 1$. Equation \eqref{eq:cv} is sometimes referred to as the \emph{white noise acceleration model}.
The initial velocity for each target is sampled from $\calN(0, 5^2)$. At every time step each target may die with probability $p_{\text{death}} = 0.05$ and new targets enter at an average rate $\lambda_{birth} = 0.5w^2$.

The measurements of each sensor follow the range-bearing model \cite{Vo19-MultiSensorMultiObject} with additive Gaussian noise sampled from $\calN(0,10^2)$ meters in range and $\calN(0,0.035^2)$ radians in bearing. The maximum detection range is $R = 2$ km with detection probability $p_d = 0.95$. Finally, the positions of clutter events are uniformly distributed within the sensor detection radius and their arrival is modeled by a Poisson distribution with $\lambda_C = 40$ events per sensor per second.

To train the model we generate a dataset of input-output image pairs \( \bfU(\calZ_n), \bfV(\calX_n) \), which are sampled from a \(w = 1\)km window with a resolution of $\rho = \frac{128}{1000}$ meters per pixel. We ran 10,000 simulations of the targets and sensors with 100 time steps each. At each time step, we advance the target dynamics following \eqref{eq:cv} and record their states \(\calX_n\). Then we simulate detections \( \calZ_n \) according to the range-bearing model. We train the model using AdamW with a batch size of 64 and a learning rate of \( 6.11 \times 10^{-6} \) for 84 epochs.

We follow a similar prodecure to generate a test dataset. At each window size \(w \in \{1,2,3,4,5\} \)km we run 100 simulations with 100 time steps each. Figure \ref{fig:multiscale} shows example inputs and outputs of the CNN, picked at random during testing at each window size. We observe that the proposed approach is able to accurately detect targets in all settings. It is noteworthy that although the CNN is trained on $1 \text{km} \times 1 \text{km}$ areas it can be successfully executed on $5 \text{km} \times 5 \text{km}$ areas. To quantify the CNN performance we extract a set \( \hat\calX_n \) of estimated target positions from the output image. We use the L1 norm of the image to determine the cardinality of this set. Then we use the k-means algorithm \cite{Pedregosa11-Scikitlearn} to find peaks in the image.

\begin{figure}[!h]
    \centering
    \includegraphics[width=\linewidth]{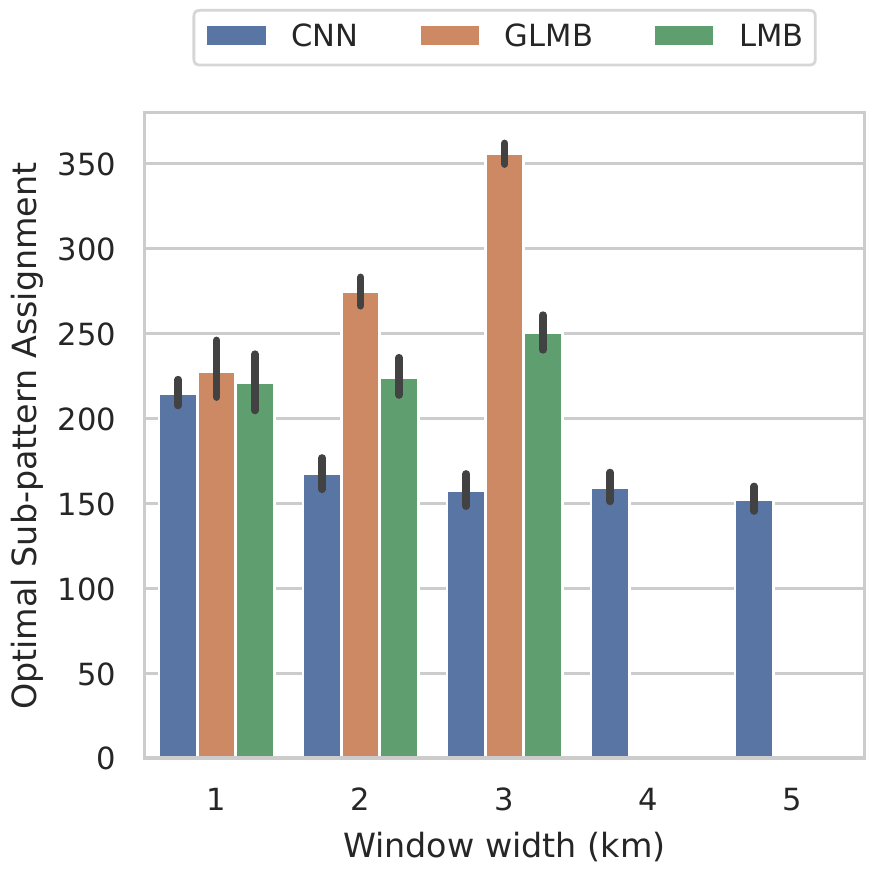}
    \caption{Comparison of OSPA of the three filters for different window sizes \(w\). We report the mean value for 100 simulations at each scale. Error bars indicate a 0.95\% confidence interval.}
    \label{fig:ospa_transfer}
\end{figure}

We benchmark the performance of the proposed CNN architecture with two state-of-the-art filters: the LMB \cite{Reuter2014, Reuter2017} and the GLMB \cite{Papi16-IteratedCorrectorGLMB}. We use the following parameters: a maximum birth probability of 1.0, an expected birth rate of 1.0, and a maximum association probability of 0.001. To implement both filters, we used the Monte Carlo approximation for the adaptive birth procedure from \cite{Trezza22-MultisensorAdaptiveBirth}. We use \(S_\text{birth} = 2000, 4000, 6000 \) Gibbs iterations for width \(w = 1, 2, 3\)km, respectivelly. The number of iterations increases because the number of sensor measurement increases.

Our results show that the performance of the CNN improves as we scale up. This is clearly shown by Figure \ref{fig:ospa_transfer}. We quantify the tracking performance of each approach using Optimal Sub-Pattern Assignment (OSPA) \cite{Schuhmacher08-Consistent} with a cutoff at \(c = 500\) meters and order \(p = 2\). OSPA measures the average distance in meters between the target positions and their estimates. The CNN achieves an OSPA of \( 215 \pm 73 \)m for a 1km\textsuperscript{2} window, which decreases to \( 152 \pm 39 \)m for a 25km\textsuperscript{2} window. This constitutes a \(29\%\) decrease in the average distance between the CNN estimates and target positions. In contrast, the performance of the RFS approaches worsens by \(57\%\) and \(13\%\) for the GLMB and LMB filters, respectively. Additionally, the CNN filter outperforms the RFS filters at every scale. For a 1km\textsuperscript{2} window, this is a marginal improvement of 6.0\% and 2.8\% compared to the GLMB and LMB filters, but the gap widens as we scale up.

\section{Conclusions}
In this paper, we studied the problem of multi-target tracking (MTT) from a deep learning perspective. In particular, we model the actual target positions and the noisy sensor measurements as multi-dimensional signals and train a convolutional neural network (CNN) to learn the MTT task. Our analysis demonstrates on the transferability of CNNs from small to arbitraly large tasks. Experimental results substantiate that the proposed transfer learning approach is a scalable solution to MTT. This offers a new perspective on a classical signal processing problem and is applicable to other domains.

% References should be produced using the bibtex program from suitable
% BiBTeX files (here: strings, refs, manuals). The IEEEbib.bst bibliography
% style file from IEEE produces unsorted bibliography list.
% -------------------------------------------------------------------------
\clearpage
\bibliographystyle{style/IEEEbib}
\bibliography{bib/icassp,bib/lmco}

\end{document}

%% file: figures/architecture_small.tex
\begin{tikzpicture}
    \pgfdeclarelayer{bg}    % declare background layer
    \pgfsetlayers{bg, main} % set layer order (main is the standard layer)

    % Control separation between blocks
    \def \deltainput     {(0.0, 0.6)}
    \def \deltaoutput    {(0.0,-0.7)}
    \def \deltalayer     {1.5}
    \def \xlabel         {3.95}
    \def \deltasigma     {(6.1, 0.0)}
    \def \deltapad       {-0.3, 1.05}

    % compute y position of Lth layer
    \newcommand{\ylayer}[1]{-#1*\deltalayer}

    % Definition of a block for block diagrams
    \tikzstyle{block} = [
    rectangle,
    minimum width = 3.0cm,
    minimum height = 1.0cm,
    font = \small,
    ]

    \tikzstyle{wide} = [
    block,
    minimum width = 6.3cm
    ]

    \tikzstyle{blockstyle} = [
    fill opacity = 0.7,
    text opacity = 1.0,
    draw = black,
    text = black
    ]

    \tikzstyle{boundingbox} = [
    dashed
    ]

    \tikzstyle{conv} = [blockstyle, fill = pastel0]
    \tikzstyle{convlocal} = [blockstyle, fill = pastel1]
    \tikzstyle{downsample} = [blockstyle, fill = pastel3]
    \tikzstyle{sigma} = [blockstyle, fill = pastel4]
    \tikzstyle{upsample} = [blockstyle, fill = pastel2]

    % Text displayed within different kinds of blocks
    \newcommand{\layerconv}[1]{$\displaystyle{
                \bfZ_{#1}[i,j] = \sum_{m_1,m_2} \red{\bfH_{#1}[m_1,m_2]}~ \bfX_{\pgfinteval{#1-1}}[i-m_1,j-m_2]
            }$}
    \newcommand{\layerconvlocal}[1]{$\displaystyle{
                \bfZ_{#1}[\bfn] = \red{\bfH_{#1}} \bfX_{\pgfinteval{#1-1}}[\bfn]
            }$}
    \newcommand{\layerdownsample}[1]{$\displaystyle{
                \bfX_{#1}[\bfn] = \sigma \Big( \bfZ_{#1}[M\bfn] \Big)
            }$}
    \newcommand{\layersigma}[1]{$\displaystyle{
                \bfX_{#1}[\bfn] = \sigma \Big( \bfZ_{#1}[\bfn] \Big)
            }$}
    \newcommand{\layerupsample}[1]{$\displaystyle{
                \bfX_{#1}[\bfn] = \sigma \Big( \bfZ_{#1}[\bfn/M] \Big)
            }$}

    % Encoder Layers
    \foreach \L in {1}
        {
            %
            % Draw Connectors between layers
            \path (0, \ylayer{\L}) node [wide, conv] (Filter\L) {\layerconv{\L}};
            %
            % Draw Nonlinearity Block
            \path (Filter\L) + \deltasigma node [block, downsample] (Sigma\L) {\layerdownsample{\L}};
        }
    % Hidden Layers
    \foreach \L in {3}
        {
            %
            % Draw Connectors between layers
            \path (0, \ylayer{\L}) node [wide, conv] (Filter\L) {\layerconv{2}};
            %
            % Draw Nonlinearity Block
            \path (Filter\L) + \deltasigma node [block, sigma] (Sigma\L) {\layersigma{2}};
        }
    % Decoder Layers
    \foreach \L in {5}
        {
            % Draw Connectors between layers
            \path (0, \ylayer{\L}) node [wide, conv] (Filter\L) {\layerconv{3}};
            %
            % Draw Nonlinearity Block
            \path (Filter\L) + \deltasigma node [block, upsample] (Sigma\L) {\layerupsample{3}};
        }
    % Draw connectors inside of layers
    \foreach \L in {1}
        {
            \node (Z\L) at (\xlabel, \ylayer{\L}) {$\bfZ_\L$};
            \draw [-stealth] (Filter\L.east)
            -- (Z\L)
            -- (Sigma\L.west);
        }
    \foreach \L in {3}
        {
            \node (Z\L) at (\xlabel, \ylayer{\L}) {$\bfZ_2$};
            \draw [-stealth] (Filter\L.east)
            -- (Z\L)
            -- (Sigma\L.west);
        }
    \foreach \L in {5}
        {
            \node (Z\L) at (\xlabel, \ylayer{\L}) {$\bfZ_3$};
            \draw [-stealth] (Filter\L.east)
            -- (Z\L)
            -- (Sigma\L.west);
        }

    % Connectors between layers
  
            % \Lplus = \L + 1
            % \ymid is the midpoint between two layer heights
            \pgfmathsetmacro{\ymid}{\ylayer{1}/2+\ylayer{3}/2}

            \node (Aux1) at (\xlabel, \ymid) {$\bfx_1$};
            \draw[-stealth] (Sigma1.south)
            |- (Aux1)
            -| (Filter3.north);

     \pgfmathsetmacro{\ymid}{\ylayer{3}/2+\ylayer{5}/2}

            \node (Aux1) at (\xlabel, \ymid) {$\bfx_2$};
            \draw[-stealth] (Sigma3.south)
            |- (Aux1)
            -| (Filter5.north);

    % Model input
    \draw[stealth-] (Filter1.north)
    -- ++\deltainput node[anchor=south] {$\bfX_0 \defeq \tilde{\bfV}$};

    % Model output
    \draw[-stealth] (Sigma5.south)
    -- ++\deltaoutput node[anchor=north] {\( \bm\Phi(\tilde{\bfV}; \red{\calH}) \)};

    % Light shading to signify layer groups. Layer 1   

    \begin{pgfonlayer}{bg}
        \draw[boundingbox] ($ (Filter1.west) + (\deltapad) $) rectangle ($ (Sigma1.east)  - (\deltapad) $);
        \node at ($ 0.5*(Z1.north) + 0.5*(Z1.north)+ (0,0.6)$) {\textsc{Encoder Layer}};

        \draw[boundingbox] ($ (Filter3.west) + (\deltapad) $) rectangle ($ (Sigma3.east)  - (\deltapad) $);
        \node at ($ 0.5*(Z3.north) + 0.5*(Z3.north) + (0,0.6)$) {\textsc{Hidden Layer}};

        \draw[boundingbox] ($ (Filter5.west) + (\deltapad) $) rectangle ($ (Sigma5.east)  - (\deltapad) $);
        \node at ($ 0.5*(Z5.north) + 0.5*(Z5.north)+ (0,0.6) $) {\textsc{Decoder Layer}};
    \end{pgfonlayer}
\end{tikzpicture}

%% file: camsap.bbl
\begin{thebibliography}{10}

\bibitem{Vo15-MultitargetTracking}
Ba-ngu Vo, Mahendra Mallick, Yaakov {Bar-shalom}, Stefano Coraluppi, Richard
  Osborne~III, Ronald Mahler, and Ba-tuong Vo,
\newblock ``Multitarget {{Tracking}},''
\newblock in {\em Wiley {{Encyclopedia}} of {{Electrical}} and {{Electronics
  Engineering}}}, pp. 1--15. {John Wiley \& Sons, Ltd}, 2015.

\bibitem{bar1974extension}
Y~Bar-Shalom,
\newblock ``Extension of the probabilistic data association filter to
  multi-target tracking,''
\newblock in {\em Proc. of the 5th Symp. Nonlinear Estimation, San Diego, Sept.
  1974}, 1974.

\bibitem{reid1979algorithm}
Donald Reid,
\newblock ``An algorithm for tracking multiple targets,''
\newblock {\em IEEE transactions on Automatic Control}, vol. 24, no. 6, pp.
  843--854, 1979.

\bibitem{hu2012cloud}
Guoqiang Hu, Wee~Peng Tay, and Yonggang Wen,
\newblock ``Cloud robotics: architecture, challenges and applications,''
\newblock {\em IEEE network}, vol. 26, no. 3, pp. 21--28, 2012.

\bibitem{ferri2017cooperative}
Gabriele Ferri*, Andrea Munaf{\`o}*, Alessandra Tesei, Paolo Braca, Florian
  Meyer, Konstantinos Pelekanakis, Roberto Petroccia, Jo{\~a}o Alves,
  Christopher Strode, and Kevin LePage,
\newblock ``Cooperative robotic networks for underwater surveillance: an
  overview,''
\newblock {\em IET Radar, Sonar \& Navigation}, vol. 11, no. 12, pp.
  1740--1761, 2017.

\bibitem{genovesio2006multiple}
Auguste Genovesio, Tim Liedl, Valentina Emiliani, Wolfgang~J Parak, Mait{\'e}
  Coppey-Moisan, and J-C Olivo-Marin,
\newblock ``Multiple particle tracking in 3-d+ t microscopy: method and
  application to the tracking of endocytosed quantum dots,''
\newblock {\em IEEE Transactions on Image Processing}, vol. 15, no. 5, pp.
  1062--1070, 2006.

\bibitem{mavska2014benchmark}
Martin Ma{\v{s}}ka, Vladim{\'\i}r Ulman, David Svoboda, Pavel Matula, Petr
  Matula, Cristina Ederra, Ainhoa Urbiola, Tom{\'a}s Espa{\~n}a, Subramanian
  Venkatesan, Deepak~MW Balak, et~al.,
\newblock ``A benchmark for comparison of cell tracking algorithms,''
\newblock {\em Bioinformatics}, vol. 30, no. 11, pp. 1609--1617, 2014.

\bibitem{patole2017automotive}
Sujeet~Milind Patole, Murat Torlak, Dan Wang, and Murtaza Ali,
\newblock ``Automotive radars: A review of signal processing techniques,''
\newblock {\em IEEE Signal Processing Magazine}, vol. 34, no. 2, pp. 22--35,
  2017.

\bibitem{Mahler07-Statistical}
Ronald P.~S. Mahler,
\newblock {\em Statistical Multisource-Multitarget Information Fusion},
\newblock Artech {{House}} Information Warfare Library. {Artech House},
  {Boston}, 2007.

\bibitem{Mahler14-AdvancesStatistical}
Ronald P.~S. Mahler,
\newblock {\em Advances in Statistical Multisource-Multitarget Information
  Fusion},
\newblock Artech {{House}} Electronic Warfare Library. {Artech House},
  {Boston}, 2014.

\bibitem{Vo06-GaussianMixture}
B.-N. Vo and W.-K. Ma,
\newblock ``The {{Gaussian Mixture Probability Hypothesis Density Filter}},''
\newblock {\em IEEE Transactions on Signal Processing}, vol. 54, no. 11, pp.
  4091--4104, Nov. 2006.

\bibitem{vo2003sequential}
Ba-Ngu Vo, Sumeetpal Singh, Arnaud Doucet, et~al.,
\newblock ``Sequential monte carlo implementation of the {PHD} filter for
  multi-target tracking,''
\newblock in {\em Proc. Int’l Conf. on Information Fusion}, 2003, pp.
  792--799.

\bibitem{vo2007analytic}
Ba-Tuong Vo, Ba-Ngu Vo, and Antonio Cantoni,
\newblock ``Analytic implementations of the cardinalized probability hypothesis
  density filter,''
\newblock {\em IEEE transactions on signal processing}, vol. 55, no. 7, pp.
  3553--3567, 2007.

\bibitem{Vo13-LabeledRandomFinite}
Ba-Tuong Vo and Ba-Ngu Vo,
\newblock ``Labeled {{Random Finite Sets}} and {{Multi-Object Conjugate
  Priors}},''
\newblock {\em IEEE Transactions on Signal Processing}, vol. 61, no. 13, pp.
  3460--3475, July 2013.

\bibitem{do2019tracking}
Cong-Thanh Do and Hoa Van~Nguyen,
\newblock ``Tracking multiple targets from multistatic doppler radar with
  unknown probability of detection,''
\newblock {\em Sensors}, vol. 19, no. 7, pp. 1672, 2019.

\bibitem{Reuter2014}
Stephan Reuter, Ba-Tuong Vo, Ba-Ngu Vo, and Klaus Dietmayer,
\newblock ``The labeled multi-{B}ernoulli filter,''
\newblock {\em IEEE Trans. Signal Process.}, vol. 62, no. 12, pp. 3246--3260,
  2014.

\bibitem{Vo17-EfficientImplementation}
Ba-Ngu Vo, Ba-Tuong Vo, and Hung~Gia Hoang,
\newblock ``An {{Efficient Implementation}} of the {{Generalized Labeled
  Multi-Bernoulli Filter}},''
\newblock {\em IEEE Transactions on Signal Processing}, vol. 65, no. 8, pp.
  1975--1987, Apr. 2017.

\bibitem{Vo19-MultiSensorMultiObject}
Ba-Ngu Vo, Ba-Tuong Vo, and Michael Beard,
\newblock ``Multi-{{Sensor Multi-Object Tracking With}} the {{Generalized
  Labeled Multi-Bernoulli Filter}},''
\newblock {\em IEEE Transactions on Signal Processing}, vol. 67, no. 23, pp.
  5952--5967, Dec. 2019.

\bibitem{beard2020solution}
Michael Beard, Ba~Tuong Vo, and Ba-Ngu Vo,
\newblock ``A solution for large-scale multi-object tracking,''
\newblock {\em IEEE Transactions on Signal Processing}, vol. 68, pp.
  2754--2769, 2020.

\bibitem{Pinto21-NextGeneration}
Juliano Pinto, Georg Hess, William Ljungbergh, Yuxuan Xia, Lennart Svensson,
  and Henk Wymeersch,
\newblock ``Next {{Generation Multitarget Trackers}}: {{Random Finite Set
  Methods}} vs {{Transformer-based Deep Learning}},'' June 2021.

\bibitem{Pinto22-CanDeepLearning}
Juliano Pinto, Georg Hess, William Ljungbergh, Yuxuan Xia, Henk Wymeersch, and
  Lennart Svensson,
\newblock ``Can {{Deep Learning}} be {{Applied}} to {{Model-Based Multi-Object
  Tracking}}?,'' Feb. 2022.

\bibitem{Vaswani17-AttentionAllYou}
Ashish Vaswani, Noam Shazeer, Niki Parmar, Jakob Uszkoreit, Llion Jones,
  Aidan~N Gomez, {\L}ukasz Kaiser, and Illia Polosukhin,
\newblock ``Attention is {{All}} you {{Need}},''
\newblock in {\em Adv. {{Neural Inf}}. {{Process}}. {{Syst}}.} 2017, vol.~30,
  {Curran Associates, Inc.}

\bibitem{cohen2016group}
Taco Cohen and Max Welling,
\newblock ``Group equivariant convolutional networks,''
\newblock in {\em International conference on machine learning}. PMLR, 2016,
  pp. 2990--2999.

\bibitem{RongLi03-SurveyManeuvering}
X.~Rong~Li and V.P. Jilkov,
\newblock ``Survey of maneuvering target tracking. {{Part I}}. {{Dynamic}}
  models,''
\newblock {\em IEEE Transactions on Aerospace and Electronic Systems}, vol. 39,
  no. 4, pp. 1333--1364, Oct. 2003.

\bibitem{Pedregosa11-Scikitlearn}
Fabian Pedregosa, Ga{\"e}l Varoquaux, Alexandre Gramfort, Vincent Michel,
  Bertrand Thirion, Olivier Grisel, Mathieu Blondel, Peter Prettenhofer, Ron
  Weiss, Vincent Dubourg, Jake Vanderplas, Alexandre Passos, David Cournapeau,
  Matthieu Brucher, Matthieu Perrot, and {\'E}douard Duchesnay,
\newblock ``Scikit-learn: {{Machine Learning}} in {{Python}},''
\newblock {\em Journal of Machine Learning Research}, vol. 12, no. 85, pp.
  2825--2830, 2011.

\bibitem{Reuter2017}
Stephan Reuter, Andreas Danzer, Manuel St{\"u}bler, Alexander Scheel, and Karl
  Granstr{\"o}m,
\newblock ``A fast implementation of the labeled multi-{B}ernoulli filter using
  {G}ibbs sampling,''
\newblock in {\em Proc. IEEE Intell. Veh. Symp.}, 2017, pp. 765--772.

\bibitem{Papi16-IteratedCorrectorGLMB}
Francesco Papi,
\newblock ``Multi-sensor $\delta$-{GLMB} filter for multi-target tracking using
  doppler only measurements,''
\newblock in {\em Proc. IEEE Eur. Intell. and Secur. Inform. Conf.}, 2015, pp.
  83--89.

\bibitem{Trezza22-MultisensorAdaptiveBirth}
A.~Trezza, Donald~J. Bucci, and P.~K. Varshney,
\newblock ``Multi-sensor joint adaptive birth sampler for labeled random finite
  set tracking,''
\newblock {\em IEEE Trans. Signal Process.}, vol. 70, pp. 1010--1025, Feb.
  2022.

\bibitem{Schuhmacher08-Consistent}
Dominic Schuhmacher, Ba-Tuong Vo, and Ba-Ngu Vo,
\newblock ``A {{Consistent Metric}} for {{Performance Evaluation}} of
  {{Multi-Object Filters}},''
\newblock {\em IEEE Transactions on Signal Processing}, vol. 56, no. 8, pp.
  3447--3457, Aug. 2008.

\end{thebibliography}
